\documentclass[aps,prx,twocolumn,amsmath,amssymb,showpacs,superscriptaddress,notitlepage]{revtex4-2}
\usepackage{amsmath,amssymb,amsfonts,bm}
\usepackage{graphicx}
\usepackage{dcolumn}
\usepackage{mathrsfs}
\usepackage{bbold}
\usepackage{dsfont}
\usepackage{dcolumn}
\usepackage{epstopdf}
\usepackage[colorlinks=true,linkcolor=blue,citecolor=blue, urlcolor=blue,bookmarks=false]{hyperref}
\usepackage{changes}
\usepackage{textgreek}
\usepackage{physics}  

\usepackage{algpseudocode}
\usepackage{amsmath}

\begin{document}

\title{Numerical calculation of the $k$-space second Chern number in four dimensions}

\author{Xiang Liu}\email{These authors contributed equally to this work.}
\affiliation{Department of Physics, Hubei University, Wuhan 430062, China}

\author{Xiao-Xia Yi}\email{These authors contributed equally to this work.}
\affiliation{Department of Fundamental Subjects, Wuchang Shouyi University, Wuhan 430064, China}

\author{Zheng-Rong Liu}
\affiliation{Department of Physics, Hubei University, Wuhan 430062, China}

\author{Rui Chen}\email{chenr@hubu.edu.cn}
\affiliation{Department of Physics, Hubei University, Wuhan 430062, China}

\author{Bin Zhou}\email{binzhou@hubu.edu.cn}
\affiliation{Department of Physics, Hubei University, Wuhan 430062, China}
\affiliation{Key Laboratory of Intelligent Sensing System and Security of Ministry of Education, Hubei University, Wuhan 430062, China}
\affiliation{Wuhan Institute of Quantum Technology, Wuhan 430206, China}

\begin{abstract}
We propose a method to numerically calculate the $k$-space second Chern number in four-dimensional (4D) topological systems. We employ an adaptive mesh refinement method to evaluate the Brillouin-zone integral, which automatically increases the grid density in regions where the Berry curvature is sharply peaked. We compare our method with the 4D lattice-gauge extension of the Fukui-Hatsugai-Suzuki method and a direct uniform grid integration scheme. Compared with these approaches, our method (i) achieves the same accuracy with substantially fewer diagonalizations, and is more computationally efficient; (ii) requires minimal memory to execute, enabling calculations for larger systems; and (iii) remains accurate even near topological phase transitions where conventional methods often face challenges. These results demonstrate that the adaptive mesh refinement method is a  powerful tool for calculating the $k$-space second Chern number.
\end{abstract}
\maketitle

\section{Introduction}

In recent years, the study of topological phases has expanded from lower dimensions to higher dimensions~\cite{Chiu2016RMP}. Specifically, four-dimensional (4D) quantum Hall insulators, characterized by the second Chern number $C_2$, have attracted significant attention due to their unique properties~\cite{Sugawa2018Science,Zhang2022CPB}.  So far, 4D Chern insulators have been experimentally observed in optical lattices~\cite{PricePRL2015}, photonic systems~\cite{Zilberberg2018Nature}, acoustic lattices~\cite{ChenZG21PRX}, and electric circuits~\cite{WangY2020NC,Yu2020NSR}. Notably, the topological magnetoelectric effect in three-dimensional topological insulators provides a physical realization of the 4D quantum Hall effect, where a magnetic field variation induces a quantized electric polarization change directly governed by the second Chern number~\cite{LiYZ2026arXiv,ZhangSC2001science,Qi08PRB,Essin09prl}.

Since the second Chern number $C_2$ serves as a key topological invariant, its accurate evaluation is crucial in characterizing higher-dimensional topological phases~\cite{ZhangSC2001science}.  However, its computation is more demanding than that of the first Chern number in two dimensions~\cite{Xiao2010RMP}, as it involves the numerical integration of the Berry curvature over a 4D Brillouin zone. Recently, Mochol-Grzelak \textit{et al.} made an important
advance by extending the Fukui--Hatsugai--Suzuki (FHS) lattice-gauge formalism to
four dimensions, providing a numerical method to
compute the $k$-space $C_2$~\cite{Mochol2018QST,Fukui2005JPSP}. This framework provides a crucial foundation for the numerical exploration of 4D topological phases~\cite{ChenXD2022NSR,Adachi2025TOJ}. At the same time, its practical use in
large-scale calculations can be limited by the cost of constructing and
handling the link variables on a dense 4D
mesh, as well as by the need to sum contributions over all 4D hypercubes.
Moreover, achieving clear quantization of $C_2$ often requires repeated Hamiltonian diagonalizations across a dense $k$-point grid. In particular, near phase transition points, an exceedingly dense grid is required to approach the quantized value, which is practically very difficult.

In this work, we adopt three different numerical methods to calculate the $k$-space $C_2$. The first method is the established extended FHS method (Method I), which serves as a benchmark based on lattice gauge theory. The second is a direct uniform grid integration (Method II). This approach is faster than Method I. However, we demonstrate that it suffers from a drawback, which fails to capture the sharp Berry curvature peaks near topological phase transitions and leads to numerical divergence near phase transition points. Importantly, the third method adopts an adaptive mesh refinement scheme, which addresses this instability without sacrificing efficiency (Method III). Our comparative analysis reveals the advantage of Method III. When targeting a high precision of $\Delta C_2 \sim 10^{-3}$, Method III not only reduces the computational cost by two orders of magnitude compared to  Method I, but also offers better convergence over Method II in critical regimes. Furthermore, Method III maintains a minimal memory consumption, making it the ideal candidate for numerically calculating the $k$-space $C_2$.

\begin{figure*}[th]
	\centering
	\includegraphics[width =1.6\columnwidth]{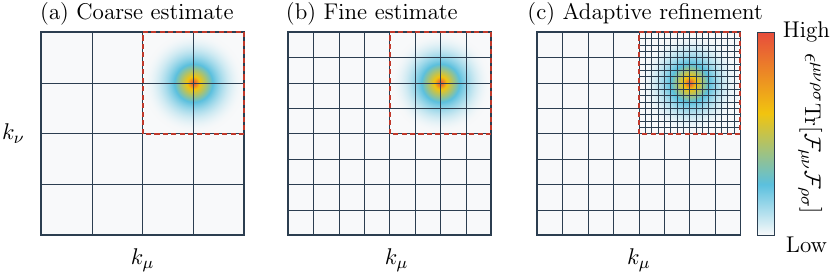}
	\caption{Schematic illustration of the adaptive mesh refinement method for evaluating the second Chern number (Method III). Here, the color denotes the distribution of  $\epsilon^{\mu\nu\rho\sigma} \text{Tr} [\mathcal{F}_{\mu\nu} \mathcal{F}_{\rho\sigma}]$, where "High" regions colored by red indicate that the Berry curvature diverges.
	(a) Coarse estimate: The integral is estimated by evaluating the density of the integration at the grid points.
	(b) Fine estimate: The hypercube is subdivided into 16 sub-cells to obtain a more precise integral value.  By comparing (a) and (b), a local error indicator is obtained for each grid.
	(c) Adaptive refinement: Regions exhibiting a high error are further subdivided (enclosed by the red square). This comparison process is then applied iteratively between the new sub-cells and their subsequent refinements, until the global error converges to tolerance ($10^{-3}$).  }
	\label{fig_illustration}
\end{figure*}

\section{Numerical methods for the Second Chern number}
\subsection{The second Chern number}
We start with a 4D topological insulator $H(\bm{k})$ and  $\{\ket{u_a(\bm{k})}\}_{a=1}^{N_{\rm occ}}$ corresponds to the wave function of the occupied bands.
The non-Abelian Berry connection and curvature are defined as~\cite{Xiao2010RMP}
\begin{align}
	\mathcal{A}_\mu^{ab}(\bm{k}) &= i \langle u_a(\bm{k}) | \partial_{k_\mu} u_b(\bm{k}) \rangle, \label{Eq:Berryconnection} \\
	\mathcal{F}_{\mu\nu}(\bm{k}) &= \partial_{k_\mu}\mathcal{A}_\nu - \partial_{k_\nu}\mathcal{A}_\mu - i[\mathcal{A}_\mu, \mathcal{A}_\nu],
	\label{Eq:Berrycurvature}
\end{align}
where $\mu,\nu \in \{1,2,3,4\}$. The second Chern number $C_2$ is given by the integral:
\begin{equation}
	C_2 = \frac{1}{32\pi^2} \int_{\rm BZ} d^4k \, \epsilon^{\mu\nu\rho\sigma} \text{Tr} \left[ \mathcal{F}_{\mu\nu}(\bm{k}) \mathcal{F}_{\rho\sigma}(\bm{k}) \right].
	\label{Eq:2CN}
\end{equation}
 Here, the trace is taken with respect to the occupied band index, $\epsilon^{\mu\nu\rho\sigma}$ denotes the 4D antisymmetric Levi-Civita symbol, and the Einstein summation convention is implied for repeated indices. Using the antisymmetry $\mathcal{F}_{\mu\nu}=-\mathcal{F}_{\nu\mu}$, the
integrand can be written in terms of three pairings, e.g.
\begin{equation}
	\epsilon^{\mu\nu\rho\sigma}
	\Tr\!\left[\mathcal{F}_{\mu\nu}\mathcal{F}_{\rho\sigma}\right]
	=
	8\,\Tr\!\left(
	\mathcal{F}_{12}\mathcal{F}_{34}
	+\mathcal{F}_{41}\mathcal{F}_{32}
	+\mathcal{F}_{31}\mathcal{F}_{24}
	\right).
	\label{eq:C2_pairing}
\end{equation}
In gapped phases, $C_2$ remains quantized and changes only when the bulk gap closes. We note that the main 
difficulty in numerically computing the $k$-space $C_2$ lies in evaluating this 4D 
integral.

\subsection{The three methods}
In this section, we will introduce three methods to evaluate $C_2$.  

\subsubsection{Method I: the extended FHS method}
For comparative study, we employ the lattice-gauge approach as a benchmark for calculating the $k$-space $C_2$ (Method I), which corresponds to a 4D generalization of the FHS method~\cite{Mochol2018QST,Fukui2005JPSP}.  This method is detailed in Appendix~\ref{Sec:FHSmethod}.

\subsubsection{Method II: direct uniform grid integration}
To avoid the numerical differentiation of eigenstates in Eq.~\eqref{Eq:Berryconnection}, we express the Berry connection elements in terms of the Hamiltonian derivatives~\cite{Xiao2010RMP}:
\begin{equation}
	\mathcal{A}_{\mu}^{ab}(\bm{k}) = i \frac{\langle u_a(\bm{k}) |
		\frac{\partial H}{\partial k_\mu} | u_b(\bm{k}) \rangle}{E_b(\bm{k}) -
		E_a(\bm{k}) }, \quad (a \neq b),
	\label{eq:perturbation}
\end{equation}
where $a$ and $b$ are band indices for the occupied bands. This expression allows us to evaluate the Berry connection in Eq.~\eqref{Eq:Berryconnection} and the Berry curvature in Eq.~\eqref{Eq:Berrycurvature}, without the need to construct a continuous gauge or compute numerical
derivatives of the wavefunctions. Then, we discretize the 4D Brillouin zone into a hypercubic grid with $N$
points along each dimension, with a grid spacing of $\Delta k =2\pi/N$. The integral for the second Chern number is approximated by the Riemann sum:
\begin{equation}
	C_2 \approx \frac{(\Delta k)^4}{32\pi^2} \sum_{\bm{k} \in \text{grid}}
	\epsilon^{\mu\nu\rho\sigma} \operatorname{Tr} \left[
	\mathcal{F}_{\mu\nu}(\bm{k}) \mathcal{F}_{\rho\sigma}(\bm{k}) \right].
	\label{eq:C2_Riemann}
\end{equation}
Here, the summation runs over all $N^4$ points in the
Brillouin zone and the trace is taken over the occupied-band subspace.

We note that if $|u_a(\bm{k})\rangle$ and $|u_b(\bm{k})\rangle$ 
are degenerate states within the occupied subspace, the energy 
denominator in Eq.~\eqref{eq:perturbation} vanishes. However, 
as demonstrated in Appendix~\ref{Sec:CancellationProof}, 
such intra-band terms cancel out with the commutator term 
$-i[\mathcal{A}_\mu, \mathcal{A}_\nu]$ in the definition of the 
non-Abelian Berry curvature. The final expression for the Berry 
curvature reads
\begin{equation}
	\mathcal{F}_{\mu\nu}^{ab} =
	i \sum_{\eta \in \text{unocc}}
	\left(
	\mathcal{A}_\mu^{a\eta} \mathcal{A}_\nu^{\eta b}
	-
	\mathcal{A}_\nu^{a\eta} \mathcal{A}_\mu^{\eta b}
	\right),
	\qquad a,b\in \text{occ}.
\end{equation}
where the summation index $\eta$ runs over the 
unoccupied bands ($\eta \in \text{unocc}$). Consequently, the 
evaluation of the Berry curvature depends solely on transitions 
between occupied and unoccupied bands, ensuring that intra-band 
degeneracies cause no numerical divergence as 
long as the bulk gap remains open.

From Eq.~\eqref{eq:perturbation}, we observe that the Berry curvature is inversely proportional to the energy difference $E_b(\bm{k}) - E_a(\bm{k})$. In the vicinity of a topological phase transition, the bulk gap closes, meaning the denominator approaches zero, and consequently, the Berry curvature exhibits sharp divergences in the Brillouin zone. As we will show below, a uniform grid fails to capture these localized singularities, leading to unstable and non-quantized values of $C_2$. Thus, Method II encounters difficulties near topological phase transitions.

\subsubsection{Method III: Adaptive mesh refinement method}
To capture the singularities of Berry curvature, we propose an adaptive mesh refinement scheme (Method III), as schematically illustrated in Fig.~\ref{fig_illustration}. Instead of a static grid, this method dynamically concentrates on regions where the Berry curvature varies rapidly. Specifically, the strategy employs a recursive approach. For each hypercube in the integration domain, we construct a local error estimator by comparing the integral contribution calculated at its geometric center [i.e., the coarse estimate in Fig.~\ref{fig_illustration}(a)], with a refined value obtained by subdividing the cell into sixteen sub-cells [i.e., the fine estimate in Fig.~\ref{fig_illustration}(b)].

The difference between the coarse and fine estimates serves as a local error indicator. If this difference exceeds a threshold within a given hypercube, the region is identified as having a high curvature gradient and is subsequently split into smaller daughter cells.
This process, shown as the adaptive refinement in Fig.~\ref{fig_illustration}(c), repeats iteratively. By comparing the refined values of the sub-cells with their own subsequent subdivisions, the grid is allowed to automatically concentrate on the singularities until the global error converges to the desired tolerance. This dynamic allocation of grid points ensures both numerical stability and computational efficiency. 

\subsection{Comparison among the three methods}

Before proceeding, we compare the computational complexity and memory requirements of the three methods. 

\paragraph{Computational Complexity}
Since the diagonalization of the Hamiltonian is the most time-consuming part, we define the complexity as the total number of diagonalizations, $N_{k}$, required for convergence. For both the lattice-gauge approach (Method I) and the uniform grid integration (Method II), the complexity scales rigidly as $N_{k}=N^4$, where $N$ is the grid density for each dimension. To resolve sharp curvature peaks near phase transitions, $N$ must be large globally, leading to unnecessary computational waste in regions where the Berry curvature is flat.
In contrast, the adaptive mesh method (Method III) decouples computational cost from global grid density. By dynamically allocating resources only to diverge regions, it achieves high precision with a significantly reduced number of diagonalizations, as will be demonstrated in the subsequent numerical calculations.

\paragraph{Memory usage}
Method I involves link variables between neighboring $k$ points. To avoid redundant diagonalizations, the eigenstates of the entire grid must be stored, with the memory cost being proportional to $O(N^4)$.  Alternatively, one could choose not to store the wavefunctions to save memory. However, this would lead to a significantly higher computational cost, requiring each Hamiltonian to be re-diagonalized $2^4 = 16$ times. Conversely, Methods II and III are strictly local. Both the perturbative Berry connection formula and the adaptive integration require information only at the current $k$-space point, independent of its neighbors. Consequently, these two methods do not need global wavefunction storage, resulting in a minimal $O(1)$ memory usage.

\begin{figure*}[th]
	\centering
	\includegraphics[width =2\columnwidth]{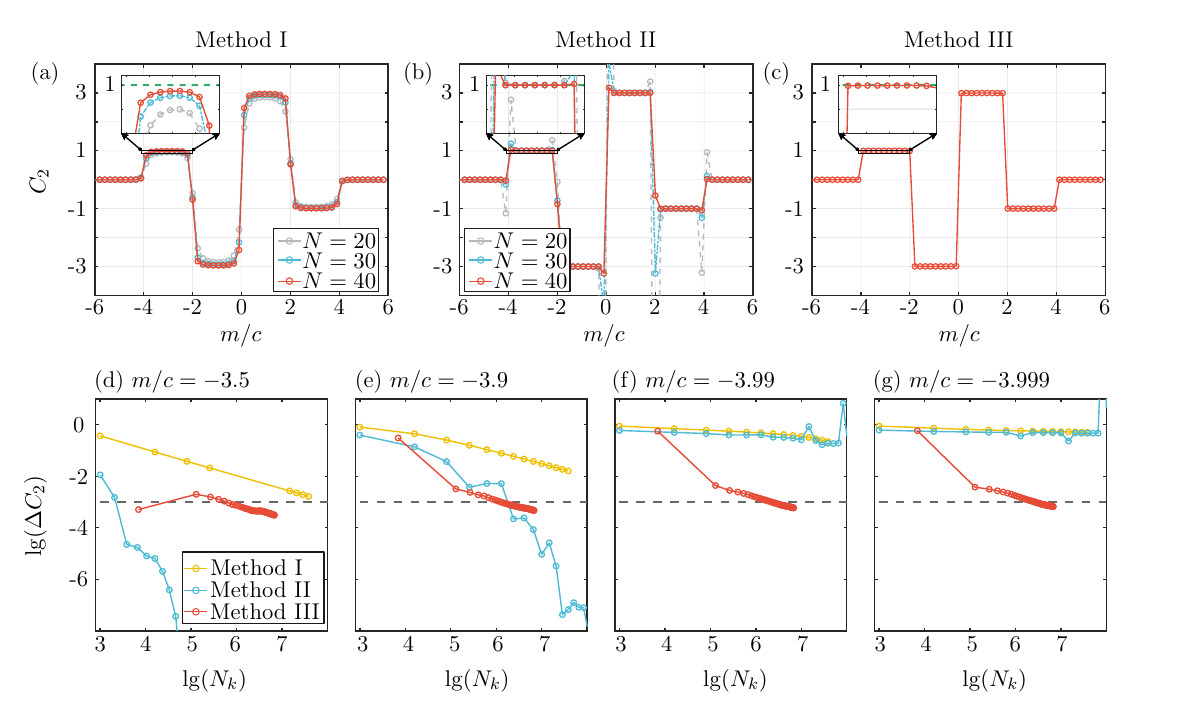}
	\caption{(a)-(c) Numerically calculated $k$-space $C_2$ as a function of $m/c$. In (a) and (b), $N_k = N^4$ denotes grid sizes. The results in (c) are obtained by setting a precision threshold of $\Delta C_2 = 10^{-3}$, where $\Delta C_2 =\left| C_{2}^{\text{calc}} - C_{2}^{\text{ideal}} \right|$ denotes the deviation of the calculated value $C_{2}^{\text{calc}}$ from the ideal theoretical value $C_{2}^{\text{ideal}}$.
	The insets provide a magnified view near the phase transition points.
	(d)-(g) Numerical error $\lg(\Delta C_2)$ as a function of the total number of diagonalizations $\lg(N_k)$ for different parameters: 
		(d) far from the transition ($m/c = -3.5$),
		(e) approaching the transition ($m/c = -3.9$),
		(f) near the critical point ($m/c = -3.99$), and
		(g) in the immediate vicinity of the transition ($m/c = -3.999$). Here,  $\lg x \equiv \log_{10} x$. The horizontal black dashed lines in (d-g) indicate a threshold of
		$\Delta C_2 = 10^{-3}$, which we consider to be sufficient  to define a quantized topological phase. }
	\label{fig_line}
\end{figure*}

\section{Results in the 4D Dirac model}
Here, we compare the performance of the three numerical methods by adopting the 4D Dirac model~\cite{Qi08PRB,Mochol2018QST},
\begin{equation}
H_{\text {latt }}=\sum_{k} \psi_{k}^{\dagger} d_a(k) \Gamma^a \psi_{k},
\end{equation}
where $d(k)$$=(m+c \sum_j \cos k_j, \sin k_1, \sin k_2, \sin k_3, \sin k_4)$, and the Dirac matrices \begin{equation}
\Gamma=\left(\sigma_x \otimes I , \sigma_y \otimes I , \sigma_z \otimes \sigma_x, \sigma_z \otimes \sigma_y, \sigma_z \otimes \sigma_z\right),
\end{equation} are defined as tensor products of the Pauli matrices $\sigma_\mu$. 

\begin{figure*}[ht]
	\centering
	\includegraphics[width =2\columnwidth]{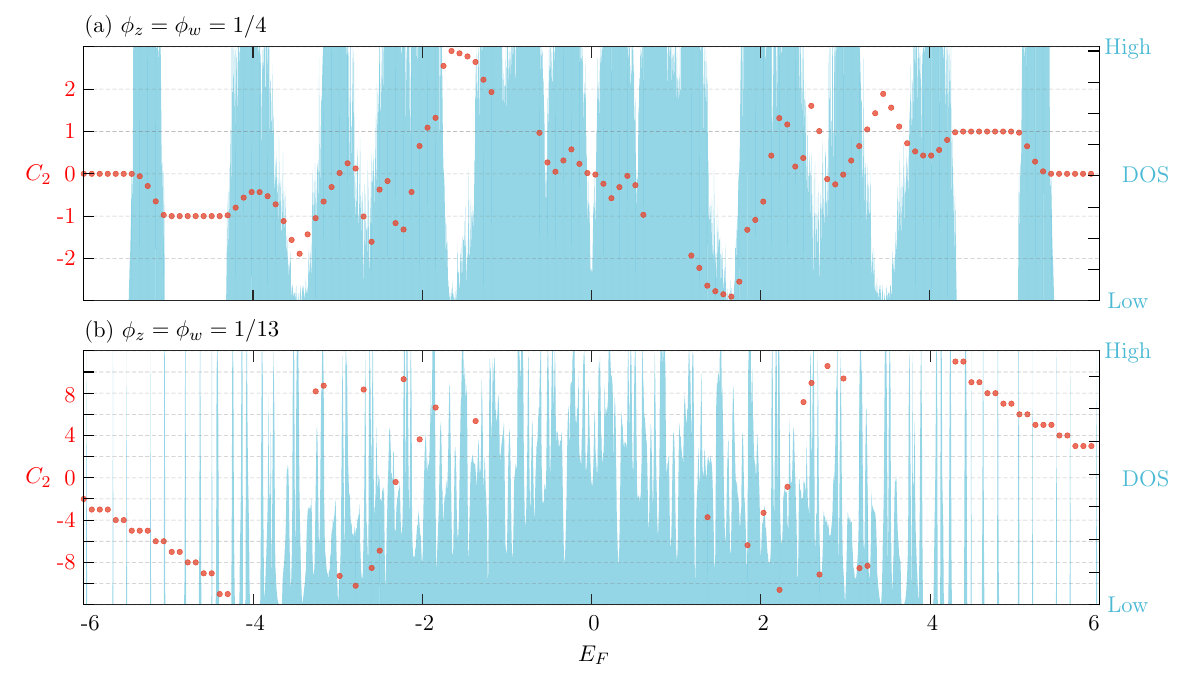}
	\caption{The evolution of $C_2$ (red) and the density of state (cyan) as a function
		of the Fermi energy $E_F$. Within the energy gaps, where the density of states vanishes, quantized plateaus of $C_2$ are observed.}
	\label{fig_magnetic}
\end{figure*}

\subsection{Accuracy and numerical stability}
Here, we focus on the accuracy of the numerically calculated $C_2$, particularly when approaching the topological phase transition points, as illustrated in Figs. \ref{fig_line}(a)-(c). Based on the numerical results, we observe distinct behaviors
regarding the precision and stability of the three methods.

Both Methods I and II achieve high precision when the system is far from the phase transition points [e.g., $m/c = -3.5$ in Figs.~\ref{fig_line}(a) and \ref{fig_line}(b)].
However, their reliability drops near the phase transition points [see the insets in Figs.~\ref{fig_line}(a) and \ref{fig_line}(b)]. Specifically, the results of Method I deviate from the quantized value. Increasing the grid density $N$ for each dimension brings the value closer to quantization, yet the accuracy is still unsatisfactory. Meanwhile, Method II exhibits fluctuations in the vicinity of the phase transition points. Method III demonstrates the best performance [Fig.~\ref{fig_line}(c)]. It yields quantized values across the
entire parameter space. Even in the immediate
vicinity of the phase transition points, Method III
maintains its accuracy where other methods fail.

\subsection{Convergence speed}
The computational efficiency of the three methods is further
elucidated by comparing their convergence rates as a function
of the number of diagonalizations $N_k$, as illustrated in
Figs.~\ref{fig_line}(d)-\ref{fig_line}(g). In the regime far from the
topological phase transition [e.g., $m/c = -3.5$ in Fig.~\ref{fig_line}(d)], Method II
exhibits the most rapid convergence, reaching a low error
threshold with minimal computational cost. Method III follows
as the second most efficient approach, while Method I
demonstrates the slowest convergence, requiring a
significantly larger $N_k$ to yield the same level of
numerical accuracy.

As the system approaches the phase transition point [$m/c = -3.9$  in Fig.~\ref{fig_line}(e)],
distinct shifts in performance emerge. The convergence rate
of Method I slows down. Although Method II still
manages to converge, it begins to exhibit numerical
instability, characterized by oscillations in the
curve. This suggests that while Method II is efficient in regions with a large topological gap with smooth Berry curvature, its reliability diminishes near the transition due to the emergence of sharp curvature singularities.

In the immediate vicinity of the critical point
[$m/c = -3.99$ and $-3.999$ in Figs.~\ref{fig_line}(f) and \ref{fig_line}(g)], both Method I and Method II
fail to provide converged topological invariants. Even with a highly dense $k$-grid involving
$N_k \approx 10^8$ diagonalizations, they fail to reach an integer value. In  contrast, the results obtained through Method III remain quantized near the topological phase transition points. It yields an integer value of $C_2$ with a precision of $\Delta C_2 \approx 0.001$
using only $10^6$ diagonalizations. This demonstrates a
clear advantage in both efficiency and reliability
for numerically calculating the $k$-space second Chern number.

\subsection{Limitations of Method III}

Based on the above results, we conclude that Method III is
the most efficient approach to calculate the $k$-space $C_2$ between the three methods. However, it is essential
to point out a specific limitation of Method III: once the precision reaches the
$10^{-3}$ threshold, the rate of convergence towards the exact theoretical value slows down.

This phenomenon is attributed to the adaptive strategy, which concentrates a higher density of grid points around the singularities of the Berry curvature. This strategy captures the singularities of the Berry curvature near topological phase transition points. However, it uses fewer points in regions where the Berry curvature is smooth, leading to minor errors there. Despite this, Method III remains the best choice for the calculation of the $k$-space $C_2$, as the precision is enough to determine the topological boundary.

Moreover, we note that the adaptive refinement scheme carries a potential risk of missing extremely narrow peaks that fall between initial sampling points. However, this issue can be reduced by choosing a sufficiently large number of initial grid points. In our implementation, the algorithm starts with $8^4$ initial hypercubes. This initial resolution has been demonstrated to be sufficient to  capture all sharp Berry curvature peaks for both the 4D Dirac system and the 4D quantum Hall system studied in this work.

\section{Application to the 4D quantum Hall system}

To further validate the robustness and versatility of Method III, we extend our investigation to a 4D quantum Hall
system described by a generalized Harper Hamiltonian~\cite{ZhangSC2001science,Kraus13PRL,PricePRL2015},
\begin{align}
	H( k )= & -J\left(e^{i k_x} u_{m+1, n}( k )+e^{-i k_x} u_{m-1, n}( k )\right) \nonumber \\
	&-J\left(e^{i k_y} u_{m, n+1}+e^{-i k_y} u_{m, n-1}\right)  \nonumber  \\
	& -2 J u_{m, n}( k )\cos \left(2 \pi \phi_z m+k_z\right)  \nonumber \\
	&-2 J u_{m, n}( k )\cos \left(2 \pi \phi_w(n+m)+k_w\right),
\end{align}
where $m$ and $n$ are the lattice indices. To construct periodic boundary conditions with rational magnetic fluxes $\phi_z=p_1 / q_1$ and $\phi_w=p_2 / q_2$, the dimension of the system (i.e., the magnetic unit cell) is set to $q_1 \times q_2$.

We first consider the case where the magnetic fluxes are $\phi_z = \phi_w = 1/4$ [Fig.~\ref{fig_magnetic}(a)]. The numerical results are calculated by using Method III, with a threshold of $10^{-3}$. The result
is consistent with the previous findings in Ref.~\cite{Mochol2018QST}.
Within the gapped regions of the energy spectrum, the quantized
second Chern number $C_2$ remains stable and clearly
defined. 

\begin{figure}[th]
	\centering
	\includegraphics[width =\columnwidth]{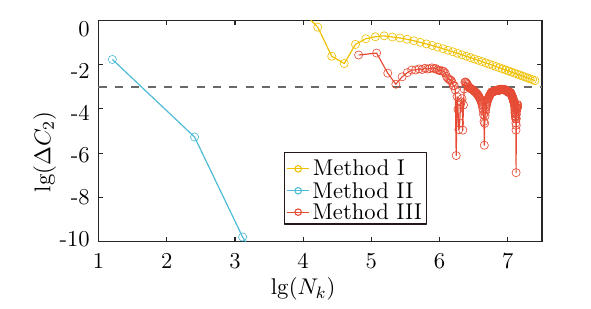}
	\caption{Convergence analysis of $C_2$ in the 4D quantum Hall system with $\phi_z = \phi_w = 1/13$ and $E_F = -4.86$. The vertical axis displays the logarithmic deviation $\lg(\Delta C_2)$,  where $\Delta C_2 =\left| C_{2}^{\text{calc}} - C_{2}^{\text{ideal}} \right|$ and $C_{2}^{\text{ideal}} = -7$. The horizontal axis $\lg(N_k)$ represents the computational cost, quantified by the total number of Hamiltonian diagonalizations. The horizontal black dashed lines indicate a threshold of
	$\Delta C_2 = 10^{-3}$, which we consider to be sufficiently quantized to define a topological phase.}
	\label{fig_magpoint}
\end{figure}

Furthermore, we explore the case with a higher magnetic
periodicity by setting $\phi_z = \phi_w = 1/13$ [Fig.~\ref{fig_magnetic}(b)]. In this regime,
the energy spectrum hosts narrower gaps, and the extended
magnetic periodicity significantly enlarges the dimension of the system. We note that applying Method I
to such a system becomes challenging, because caching the global
wave functions in such systems imposes
significant memory demands. Nevertheless, Method III successfully
captures distinct plateau-to-plateau topological phase transitions.
As the Fermi energy $E_F$ scans through the gaps, the second Chern
number exhibits well-quantized jumps between different integer values.
The results confirm that the stability of Method III is not
model-specific.

To quantitatively analyze the computational efficiency, we focus on the case of $\phi_z = \phi_w = 1/13$ at $E_F = -4.86$. At this Fermi energy, the density of states vanishes, confirming that the system lies within a bulk energy gap where the second Chern number $C_2$ is quantized. We determine the benchmark value from the gauge-invariant extended FHS calculation on sufficiently fine meshes and confirm that it corresponds to a stable quantized plateau as $E_F$ is varied within the same gap. This gives the reference value $C_2=-7$. The results obtained by the uniform-grid method and the adaptive method converge to this same integer as the number of diagonalizations is increased.

Figure~\ref{fig_magpoint} illustrates the convergence of the numerically calculated $C_2$ as a function of the number of Hamiltonian diagonalizations $N_k$. The three methods exhibit distinct behaviors: Method I shows excellent stability but suffers from the slowest convergence rate. Method II provides a significant improvement in speed over Method I. Method III demonstrates a moderate convergence rate, achieving the target precision with approximately two orders of magnitude fewer diagonalizations than Method I. Taking both stability and efficiency into account, Method III is the best choice; Method II is suitable for rapid estimations; and Method I can be employed to verify the final results.

We note that Method III exhibits a minor nonmonotonic fluctuation and a temporary slowing down of convergence around $\lg(N_k)\sim 6,7$. This behavior originates from the adaptive refinement process. After the main sharp peaks of the Berry curvature have been located and refined, the remaining error is partly controlled by the residual discretization error in the less-refined smooth regions. Therefore, a small fluctuation may appear when the mesh is redistributed among different regions of the Brillouin zone. This is a numerical feature of the adaptive procedure, rather than a physical instability or a fundamental obstacle. The accuracy can be systematically improved by increasing the number of initial grid cells before the adaptive refinement starts, which gives a better global sampling of the Brillouin zone and reduces the residual background integration error. In practice, the final result can be validated by checking convergence with respect to the initial mesh size. For the systems studied here, a resolution of $10^{-3}$ remains sufficient to determine the quantized topological boundary.
\section{Conclusion}

In conclusion, we propose an adaptive mesh refinement method (Method III) to numerically calculate the $k$-space second Chern number $C_2$ in 4D topological systems.  While the lattice-gauge formulation (Method I) and uniform grid integration (Method II) fail to capture the quantized $C_2$ near topological phase transition points, Method III overcomes this limitation, maintaining high precision and robust quantization in the vicinity of critical points. Furthermore, for a given accuracy requirement, Method III operates significantly faster than Method I and has negligible memory requirements. Therefore, Method III serves as an efficient and reliable tool for calculating the $k$-space $C_2$ in 4D topological systems.

Furthermore, we emphasize that the applicability of the adaptive mesh refinement method extends beyond the second Chern number. It is also applicable to the evaluation of different physical observable or topological invariant that requires the integration of geometric quantities over the Brillouin zone, such as the third Chern number in six dimensions~\cite{Petrides2018PRB}, and the nonlinear Hall effect associated with the Berry curvature dipole~\cite{Sodemann15prl,Du2021NRP} or quantum metric~\cite{WangC21PRL,Liu21PRL}. We will systematically explore these broader applications in our future work.

In two dimensions, the computational burden of a uniform mesh scales as $N^2$, much more mildly than the $N^4$ scaling in the four-dimensional second-Chern-number problem. Moreover, the original FHS method in two dimensions is already very efficient, gauge invariant, and robust for many standard models. Therefore, for simple two-dimensional models far from a phase transition, our adaptive strategy may not provide a clear advantage over the FHS method. The adaptive idea can still be useful in two dimensions when the Berry curvature is strongly localized, for example near a gap closing or in large Hilbert-space models where each diagonalization is expensive. In that case, it can concentrate sampling points near the sharp curvature region.

We also note that the efficiency of Method III is tied to the number of Berry curvature peaks. If a system exhibits a very large number of sharp peaks widely distributed across the Brillouin zone, the frequent cell-splitting process will introduce additional computational overhead, thereby reducing the efficiency of the adaptive scheme. Nevertheless, compared to a globally dense uniform grid, the method can still reduce the total number of diagonalizations to some extent by maintaining a coarser mesh in the remaining smooth regions.
	
\appendix
\section{Details of the extended FHS method}
\label{Sec:FHSmethod}
The link tensor describes the phase acquired
during the parallel transport between two neighboring sites in the momentum space~\cite{Mochol2018QST,Fukui2005JPSP}. For occupied states $\psi^\alpha$, the link
variable is given by:
\begin{equation}
	\left(U_\mu\right)^{\alpha \beta}(k_l) =
	\left\langle\psi^\alpha( k_l ) \mid \psi^\beta\left( k_l +\Delta
	\hat{\mu}\right)\right\rangle,
\end{equation}
where $k_l$ denotes a discrete momentum-space point. Using these
link variables, the Wilson loop is constructed around a 
plaquette spanning directions $\mu$ and $\nu$:
\begin{equation}
	U_{\mu \nu}^P(k_l) = U_\mu^P\left( k_l\right) U_\nu^P\left(
	k_l+\hat{\mu}\right) U_\mu^P\left( k_l+\hat{\nu}\right)^{-1}
	U_\nu^P\left( k_l\right)^{-1},
\end{equation}
where $U_\mu^P( k ) \equiv P U_\mu( k ) P$ involves the projection onto
the relevant bands.

The lattice field strength (non-Abelian Berry curvature) on each
plaquette is extracted via the matrix logarithm of the Wilson loop:
\begin{equation}
	\tilde{F}_{\mu \nu}^P\left( k_l\right) = \ln \left(U_{\mu
		\nu}^P(k_l)\right).
\end{equation}
To evaluate the second Chern number, we compute the contribution from
each hypercube cell of the hyperlattice. The total invariant
$\tilde{C}_2$ is obtained by summing the trace of the symmetrized field
strength products over all grid points $\{k_l\}$:
\begin{equation}
	\tilde{C}_2\left(\varepsilon_F\right) = \frac{1}{4 \pi^2}
	\sum_{\left\{ k_l\right\}} \operatorname{Tr} \tilde{F}_l^P,
\end{equation}
where the local contribution $\tilde{F}_l^P$ is defined by the cyclic
permutation of indices:
\begin{align}
	\tilde{F}_l^P &= \tilde{F}_{x y}^P\left( k_l\right) \tilde{F}_{z
		w}^P\left( k_l\right) + \tilde{F}_{w x}^P\left( k_l\right)
	\tilde{F}_{z y}^P\left( k_l\right) \nonumber
	\\
	&+ \tilde{F}_{z x}^P\left(
	k_l\right) \tilde{F}_{y w}^P\left( k_l\right).
\end{align}
This formulation ensures that the flux is calculated in a gauge-invariant
manner on the discrete mesh, giving an integer index to capture the topological nature of the 4D system.

\section{Details of the non-Abelian Berry curvature}
\label{Sec:CancellationProof}

In this Appendix, we show
that intra-band degeneracies within the occupied subspace 
do not lead to numerical divergences in the evaluation of 
the Berry curvature.

Let $|u_a\rangle$ and $|u_b\rangle$ be the eigenstates 
belonging to the occupied subspace ($a, b \in \text{occ}$). 
The non-Abelian Berry connection within this subspace is defined 
as $\mathcal{A}_\mu^{ab} = i \langle u_a | \partial_\mu u_b \rangle$, 
and the corresponding Berry curvature is given by
\begin{equation}
	\mathcal{F}_{\mu\nu}^{ab} = \partial_\mu \mathcal{A}_\nu^{ab} 
	- \partial_\nu \mathcal{A}_\mu^{ab} 
	- i [\mathcal{A}_\mu, \mathcal{A}_\nu]^{ab}.
\end{equation}
First, we expand the linear derivative part 
$\partial_\mu \mathcal{A}_\nu^{ab}$:
\begin{equation}
	\partial_\mu \mathcal{A}_\nu^{ab} = 
	i \langle \partial_\mu u_a | \partial_\nu u_b \rangle 
	+ i \langle u_a | \partial_\mu \partial_\nu u_b \rangle.
\end{equation}
Using the fact that the partial derivatives commute 
($\partial_\mu \partial_\nu = \partial_\nu \partial_\mu$), 
the antisymmetric combination yields
\begin{equation}
	\partial_\mu \mathcal{A}_\nu^{ab} 
	- \partial_\nu \mathcal{A}_\mu^{ab} = 
	i \langle \partial_\mu u_a | \partial_\nu u_b \rangle 
	- i \langle \partial_\nu u_a | \partial_\mu u_b \rangle.
\end{equation}
Next, we insert the resolution of identity 
$\sum_{n \in \text{all}} |u_n\rangle \langle u_n| = 1$ 
over the complete set of states into the inner product:
\begin{equation}
	\langle \partial_\mu u_a | \partial_\nu u_b \rangle = 
	\sum_{n \in \text{all}} \langle \partial_\mu u_a | u_n \rangle 
	\langle u_n | \partial_\nu u_b \rangle.
\end{equation}
Differentiating the orthonormality condition 
$\langle u_a | u_n \rangle = \delta_{an}$ gives 
$\langle \partial_\mu u_a | u_n \rangle = 
- \langle u_a | \partial_\mu u_n \rangle = 
i \mathcal{A}_\mu^{an}$. Meanwhile, by definition, 
$\langle u_n | \partial_\nu u_b \rangle = 
-i \mathcal{A}_\nu^{nb}$. Substituting these into the 
summation leads to
\begin{equation}
	\langle \partial_\mu u_a | \partial_\nu u_b \rangle = 
	\sum_{n \in \text{all}} (i \mathcal{A}_\mu^{an}) 
	(-i \mathcal{A}_\nu^{nb}) = 
	\sum_{n \in \text{all}} \mathcal{A}_\mu^{an} 
	\mathcal{A}_\nu^{nb}.
\end{equation}
Thus, the linear derivative part becomes
\begin{align}
	\partial_\mu \mathcal{A}_\nu^{ab} 
	- \partial_\nu \mathcal{A}_\mu^{ab} &= 
	i \sum_{n \in \text{all}} \left( \mathcal{A}_\mu^{an} 
	\mathcal{A}_\nu^{nb} - \mathcal{A}_\nu^{an} 
	\mathcal{A}_\mu^{nb} \right) \nonumber \\
	&= i \sum_{c \in \text{occ}} \left( \mathcal{A}_\mu^{ac} 
	\mathcal{A}_\nu^{cb} - \mathcal{A}_\nu^{ac} 
	\mathcal{A}_\mu^{cb} \right) \nonumber \\
	&+ i \sum_{\eta \in \text{unocc}} \left( 
	\mathcal{A}_\mu^{a\eta} \mathcal{A}_\nu^{\eta b} 
	- \mathcal{A}_\nu^{a\eta} \mathcal{A}_\mu^{\eta b} 
	\right).
\end{align}
On the other hand, the commutator term 
$-i [\mathcal{A}_\mu, \mathcal{A}_\nu]^{ab}$ by definition 
only sums over the occupied subspace:
\begin{equation}
	-i [\mathcal{A}_\mu, \mathcal{A}_\nu]^{ab} = 
	-i \sum_{c \in \text{occ}} \left( \mathcal{A}_\mu^{ac} 
	\mathcal{A}_\nu^{cb} - \mathcal{A}_\nu^{ac} 
	\mathcal{A}_\mu^{cb} \right).
\end{equation}
Combining the linear derivative part and the commutator term, 
the summation over the occupied states $\sum_{c \in \text{occ}}$ 
cancels out exactly, leaving only the contribution from the 
unoccupied states~\cite{Zhang2022CPB}:
\begin{equation}
	\mathcal{F}_{\mu\nu}^{ab} =
	i \sum_{\eta \in \text{unocc}}
	\left(
	\mathcal{A}_\mu^{a\eta} \mathcal{A}_\nu^{\eta b}
	-
	\mathcal{A}_\nu^{a\eta} \mathcal{A}_\mu^{\eta b}
	\right),
	\qquad a,b\in \text{occ}.
\end{equation}
When we express the remaining connection elements via the 
perturbation formula, $\mathcal{A}_\mu^{a\eta} = 
i \langle u_a | \partial_\mu H | u_\eta \rangle / 
(E_\eta - E_a)$, the energy denominator always depends on 
$E_\eta - E_a$. Because $a \in \text{occ}$ and 
$\eta \in \text{unocc}$, this denominator represents the 
bulk energy gap, which is strictly non-zero for an insulator. 
Any intra-band degeneracies within the occupied subspace 
itself (i.e., $E_a = E_b$ for $a, b \in \text{occ}$) are 
completely absent from the expression, thereby 
guaranteeing that no numerical divergence occurs.

\section*{Acknowledgments}
We acknowledge the support of the NSFC (under Grant Nos.~U25D8012, 12304195, 12074107, 12504560), the Chutian Scholars Program in Hubei Province, the Hubei Provincial Natural Science Foundation (Grant No. 2025AFA081, 2022CFA012), the Guangdong Provincial Quantum Science Strategic Initiative (Grant No. GDZX2401001), the Wuhan city key R\&D program (under Grant No. 2025050602030069), the program of outstanding young and middle-aged scientific and technological innovation team of colleges and universities in Hubei Province (under Grant No. T2020001), the key project of Hubei provincial department of education (under Grant No. D20241004), and the original seed program of Hubei university.

%
%
%
\bibliographystyle{apsrev4-1-etal-title_6authors}
\bibliography{refs-transport,refs-transport1}

\end{document}